\documentclass{PoS}
\usepackage{amsmath}
\usepackage{graphics}

\title{Individual eigenvalue distributions for chGSE-chGUE crossover and determination of low-energy constants in two-color QCD+QED}

\ShortTitle{Individual eigenvalue distributions for chGSE-chGUE crossover}

\author{\speaker{Shinsuke M. Nishigaki}\thanks{
SMN is supported in part by JSPS Grants-in-Aid for Scientific Research (KAKENHI) No.\ 25400259. 
}\\
       Graduate School of Science and Engineering\\
       Shimane University, Matsue 690-8504, Japan\\
       E-mail: \email{mochizuki@riko.shimane-u.ac.jp}}
\author{Takuya Yamamoto\\
       Graduate School of Science and Engineering\\
       Shimane University, Matsue 690-8504, Japan}

\abstract{
We compute statistical distributions of individual low-lying eigenvalues
of random matrix ensembles interpolating
chiral Gaussian symplectic and unitary ensembles.
To this aim we use the Nystr\"{o}m-type discretization of Fredholm Pfaffians and 
resolvents of the dynamical Bessel kernel 
containing a single crossover parameter $\rho$.
The $\rho$-dependent distributions of the four smallest eigenvalues are then used to 
fit the Dirac spectra of modulated SU(2) lattice gauge theory,
in which the reality of the staggered SU(2) Dirac operator is weakly
violated either by the U(1) gauge field or by a constant background flux.
Combined use of individual eigenvalue distributions
is effective in reducing statistical errors in $\rho$;
its linear dependence on the imaginary chemical potential $\mu_I$ 
enables precise determination of the pseudo-scalar decay constant $F$
of the SU(2) gauge theory 
from a small lattice.
The U(1)-coupling dependence of an equivalent of $F^2 \mu_I^2$ in the SU(2)$\times$U(1) theory is also obtained.
}

\FullConference{The 32nd International Symposium on Lattice Field Theory\\
		 23-28 June, 2014\\
		 Columbia University New York, NY}

\def\beq{\begin{equation}}
\def\eeq{\end{equation}}
\def\be{\begin{equation}}
\def\ee{\end{equation}}
\def\bea{\begin{eqnarray}}
\def\eea{\end{eqnarray}}
\def\ba{\begin{array}}
\def\ea{\end{array}}
\def\<{\left\langle}
\def\>{\right\rangle}
\def\({\left(}
\def\){\right)}
\def\e{{\rm e}}
\def\tr{{\rm tr}}
\def\openone{\mathbf{1}}
\def\openzero{\mathbf{0}}
\newcommand\bs[1]{\boldsymbol{\mathit{#1}}}

\begin{document}

\section{Introduction}
Wigner-Dyson universality of the local correlation of energy levels
among various stochastic \cite{Meh04} and quantum-chaotic systems \cite{BGS84}
under well-defined conditions was established through the ten-fold classification of 
symmetric spaces of spectral $\sigma$ models \cite{Zir96},
to which the Gutzwiller trace formula also reduces \cite{MHABH09}.
This universality has in turn provided a solid and secure ground 
on which system-specific information can be decoded
by measuring deviations of spectral correlation functions from their universal forms,
or transition between two universality classes.
Prime examples of the former are the weak localization correction in 
Anderson Hamiltonians \cite{Efe97} and
the nonuniversal effect of short periodic orbits (small primes) in chaotic systems
(in the Riemann $\zeta$ zeroes \cite{BK99}).

Study on the latter ``universality crossover", initiated by Dyson \cite{Dys62},
has also come to encompass a variety of settings, 
an example being the GUE-GOE transition that appears in
a disordered ring \cite{DM91} and chaotic systems \cite{SNMB09} both under magnetic fields.
Recent years saw applications of the universality crossover in lattice QCD,
in an effort to explore the effects of the isospin chemical potential \cite{DHSS05} 
and of the finite lattice spacing in the Wilson Dirac operator \cite{DSV10}.
These studies have revealed the power of the spectral approach
in determining the pion decay constant and the Wilsonian chPT constants
from relatively small lattices.
The aim of this work is to apply this approach to the determination of low-energy constants 
in another setting, namely the two-color QCD subjected under the imaginary chemical potential
\cite{Nis12} or coupled to QED.
Our novelty is to employ the individual distributions of small Dirac eigenvalues \cite{DN01}
instead of $n$-level correlation functions, in fitting the lattice data.
Practical advantages of our method will be manifested subsequently.

\section{chGSE-chGUE crossover}
Let $\bs{A}$ and $\bs{B}$ be $N/2\times N'/2$ quaternion matrices,
represented by complex $N\times N'$ matrices as
\be
A=\sum_{\mu=0}^3 \(A^{(\mu)}_{jk}\) \otimes {\bf e}_\mu,\ \  
B=\sum_{\mu=0}^3 \(B^{(\mu)}_{jk}\) \otimes {\bf e}_\mu\ \ 
(j=1,\ldots,N/2,\ k=1,\ldots,N'/2).
\ee
Here a set of four $2\times2$ matrices ${\bf e}_\mu=(\openone_2, -i\vec{\sigma})$
spans the basis of the quaternion field $\mathbb{H}$.
Let the matrix elements belong to
$
A^{(\mu)}_{jk}\in \mathbb{R}
$
and
$
B^{(\mu)}_{jk}\in \mathbb{C},
$
so that the matrix $\bs{A}$ is quaternion-real and $\bs{B}$ is not (i.e. a generic $N\times N'$ complex matrix).
We consider
$A^{(\mu)}_{jk}$, ${\rm Re}\, B^{(\mu)}_{jk}$, and ${\rm Im}\, B^{(\mu)}_{jk}$
to be independent random variables  distributed according to
the Gaussian distributions
$\e^{-\frac12\tr\, A A^\dagger}$ and $\e^{-\tr\, B B^\dagger}$,
respectively, and introduce
an ensemble of $(N+N') \times (N+N') $ Hermitian matrices $H$ of the form
\be
H=
\left(
\ba{cc}
\openzero_{N\times N} &C\\
C^{\dagger} &\openzero_{N'\times N'}
\ea
\right),
\ \ C=\e^{-\tau}A+\sqrt{1-\e^{-2\tau}} B.
\label{H}
\ee
Here a real parameter $\tau$ plays the role of
fictitious time for the Brownian motion 
of the eigenvalues \cite{Dys62}.
This ensemble enjoys
the chiral symmetry $\{H,\gamma_5\}=0$ with $\gamma_5={\rm diag}(\openone_{N},-\openone_{N'})$,
implying that the spectrum of $H$ consists of $N$ 
$\pm$ pairs of nonzero eigenvalues and $\nu=|N'-N|$ zero eigenvalues.
The presence of $B$ violates the quaternion-reality of $C$ and the selfduality of $H$,
lifting the Kramers degeneracy of nonzero eigenvalues of $H$.
Accordingly this ensemble interpolates the two limiting cases,
chiral GSE at $\tau=0$ and chiral GUE at $\tau\to\infty$,
depending on a single parameter $\tau$.

We consider the case in which the Kramers degeneracy is weakly broken by $\tau\ll 1$.
Then the spectral density of $H$ in the large-$N$ limit
is identical to that of the chGSE ($\tau=0$), i.e. Wigner's semi-circle
$\bar{\rho}(\lambda)=\sqrt{4N-\lambda^2}/\pi$.
We magnify the vicinity of the origin of the $\lambda$ axis by introducing
unfolded variables $x_i=\lambda_i/\varDelta$
with $\varDelta=1/\bar{\rho}(0)=\pi/\sqrt{4N}$.
In order to realize a nontrivial crossover behavior, 
we take the triple-scaling limit $N, N'\to\infty, \lambda_i\to 0, \tau\to 0$
while keeping the combinations $\rho=\sqrt{\tau}/\varDelta$, $\nu=N'-N(\geq 0)$, and $x_i$ fixed finite.
Then the j.p.d.~of $N$ positive unfolded eigenvalues $P_N(x_1,\ldots,x_N)$ is expressed 
as a Pfaffian of the dynamical Bessel kernel $K(x,y)$ \cite{FNH99},
\bea
&&
P_N(x_1,\ldots,x_N)={\rm Pf}\(Z \left[K(x_i, x_j)\right]_{i,j=1}^N\),
\ 
K(x, y)=\left[
\ba{cc}
S(x,y) & I(x, y)\\
D(x,y) & S(y, x)
\ea
\right],\ 
Z=\left[
\ba{cc}
0 & 1 \\
-1 & 0
\ea
\right]\otimes \openone,
\label{Pfaffian}
\\
&&S(x,y)=\pi  \sqrt{x y} \left\{
\frac{J_{\nu}(\pi  x) y J_{\nu-1}(\pi  y)-x J_{\nu-1}(\pi  x) J_\nu(\pi  y)}{x^2-y^2}
-\frac{J_\nu(\pi x)}{2}\!\!\int_0^\pi d\upsilon\,\e^{\rho^2 (\upsilon^2-\pi^2)}J_\nu(\upsilon y)\right\}
,\nonumber\\
&&D(x,y)=
\frac{\sqrt{xy} }{2} \int_0^\pi d\upsilon\,\upsilon  \int_0^1 du\,\e^{\rho^2 \upsilon^2(1+u^2)}
\left\{J_\nu(\upsilon u x) J_\nu(\upsilon y)- J_\nu(\upsilon x) J_\nu(\upsilon u y)\right\}
,\nonumber\\
&&I(x,y)=
\frac{ \sqrt{x y}}{2}  \int_\pi^\infty d\upsilon\, \upsilon^2 \,\e^{-2 \rho^2 \upsilon^2} 
\left\{J_{\nu}(\upsilon x) y J_{\nu-1}(\upsilon y)- x J_{\nu-1}(\upsilon x) J_{\nu}(\upsilon y)\right\}.
\nonumber
\end{eqnarray}
Due to the recursion relation 
$\int_0^\infty dx_k {\rm Pf}\(Z\left[K(x_i, x_j)\right]_{i,j=1}^k \)=(N-k+1){\rm Pf}\(Z\left[K(x_i, x_j)\right]_{i,j=1}^{k-1}\),$
correlation functions of $n$ eigenvalues are given by 
\be
R_n(x_1,\ldots,x_n)=
\frac{N!}{(N-n)!}
\int_0^\infty dx_{n+1}\ldots dx_N\,P_N(x_1,\ldots,x_N)
={\rm Pf}\(Z \left[K(x_i, x_j)\right]_{i,j=1}^n\).
\label{Pfaffian2}
\ee

\section{Individual eigenvalue distributions}
The  Pfaffian forms in (\ref{Pfaffian})$\sim$(\ref{Pfaffian2})
originate from quaternion determinants (Tdet)
composed of a quaternionic kernel, $\left[\mathcal{K}(x_i, x_j)\right]_{i,j}$,
whose $\mathbb{C}$-number representative is 
the antisymmetric matrix $Z \left[{K}(x_i, x_j)\right]_{i,j}$.
Accordingly,
the probability $E_k(s)$ for an interval $[0, s]$ to contain exactly $k$ eigenvalues is also given
in terms of the Fredholm Tdet of a quaternionic integral operator $\hat{{\cal K}}_s$, 
i.e.\ the square root of the corresponding Fredholm determinant of $\hat{K}_s$ 
(i.e.~Fredholm Pfaffian of $Z\hat{K}_s$),
\be
E_k(s)=
\frac{1}{k!} (-\partial_\xi)^k \left. {\rm Det} (1-\xi \hat{K}_s)^{1/2}\right|_{\xi=1}.
\label{FredholmPfaffian}
\ee
Here $\hat{K}_s$ denotes an integral operator with the dynamical Bessel kernel $K(x,y)$ (\ref{Pfaffian})
acting on the space of two-component $L^2$-functions over the interval $[0,s]$.
First few $E_k(s)$'s are expressed as
\bea
&&E_0(s)={{\rm Det} (1-\hat{K}_s)^{1/2}},\quad
E_1(s)=E_0(s)\frac{T_1}{2},\quad
E_2(s)=E_0(s)\frac{1}{2!} \left(\frac{T_1^2}{4}-\frac{T_2}{2}\right),
\label{Ek}\\
&&E_3(s)=E_0(s)\frac{1}{3!} \left(\frac{T_1^3}{8}-\frac34 T_1 T_2+T_3\right),\quad
E_4(s)=E_0(s)\frac{1}{4!} \left(\frac{T_1^4}{16}- \frac34 T_1^2 T_2+ \frac34 T_2^2+2T_1 T_3-3T_4\right),
\nonumber
\eea
where
$T_n(s)={\rm Tr} \bigl(\hat{K}_s(I-\hat{K}_s)^{-1}\bigr)^n$
denote functional traces of the resolvents of $\hat{K}_s$.
Probability distribution $p_k(s)$ of the $k^{\rm th}$ smallest positive eigenvalue is then given as
$p_k(s)=-\partial_s \sum_{\ell=0}^{k-1} E_{\ell}(s).$
An efficient way of numerically evaluating the Fredholm determinant of
a trace-class operator $\hat{K}_s$ acting on $L^2$-functions over an interval $[0,s]$
is the Nystr\"{o}m-type discretization \cite{Bor10}
\be
{\rm Det}(1-\hat{K}_s)\simeq \det(I-\mathbf{K}_s),\ \ 
\mathbf{K}_s= \left[K(x_i,x_j) \sqrt{w_i\,w_j}\right]_{i,j=1}^m ~.
\label{Nystrom}
\ee
\begin{figure}[b]
\begin{center}
\includegraphics[bb=0 0 259 165,width=74mm]{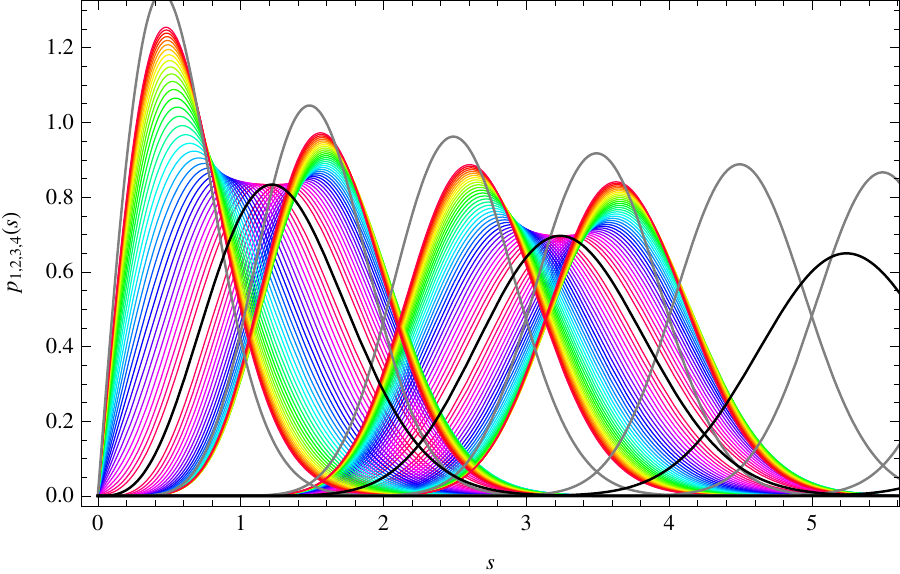}
\includegraphics[bb=0 0 258 165,width=73.3mm]{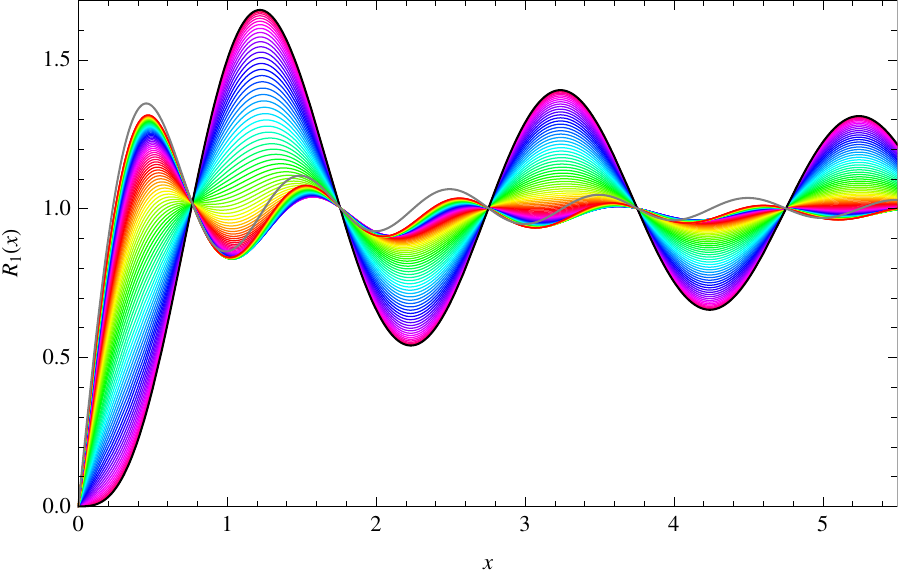}
\caption{
First four eigenvalue distributions $p_{1\sim 4}(s)$ (left) for $0.04\leq \rho \leq 0.70$ (step 0.01, purple to red)
and the spectral density $R_1(x)$ for $0.01\leq \rho \leq 1.00$ (step 0.01)
for the chGSE (black) to chGUE (grey) crossover.}
\end{center}
\end{figure}
Here we employ a quadrature rule consisting of a set of points $\{x_i\}$ 
taken from the interval $[0,s]$ and associated weights $\{w_i\}$ such that
${\int_0^s f(x)dx  \simeq \sum_{i=1}^m f(x_i) w_i}$.
Similarly, the  resolvents in (\ref{Ek}) are approximated as
$T_n(s)\simeq {\rm tr} \left(\mathbf{K}_s(I-\mathbf{K}_s)^{-1}\right)^n.$
For a practical purpose we choose the Gauss quadrature rule, i.e.\ sampling $\{x_i\}$ from
the nodes of Legendre polynomials normalized to $[0,s]$.
Previously we applied the Nystr\"{o}m-type method to the dynamical Bessel kernels
interpolating chGSE-chGUE (\ref{Pfaffian}) and chGOE-chGUE
and evaluated the smallest eigenvalue distributions $p_1(s)$ \cite{Nis12}.
In this work we extend our computation to the first four eigenvalues, aiming to reduce the fitting errors
in determining the low-energy constants.
We set the approximation order $m$ to be 
at least 100, and confirmed the stability of the results for increasing $m$ (up to $200\sim 400$).

The distributions
$p_1(s), \cdots, p_4(s)$ for the $\nu=0$ case,
computed from (\ref{Pfaffian})$\sim$(\ref{Nystrom}) 
for $\rho \leq 0.70$ are exhibited in Fig.~1L.
A practical advantage of using individual eigenvalue distributions
over the spectral density $R_1(x)=\sum_{k=1}^\infty p_k(x)=S(x,x)$ (Fig.~1R) 
for fitting the lattice data is clear from the figure:
the oscillation of the latter immediately becomes structureless and insensitive to 
the  interpolation parameter $\rho$ due to the overlapping of multiple peaks of the former, whereas
the quasi-Gaussian shape of each peak is clearly distinguishable and is extremely sensitive to $\rho$.
Another advantage specific to the current case originates from the fact that
$p_{2k-1}(s)$ and $p_{2k}(s)$ move in opposite directions as $\rho$ is increased to break the 
Kramers degeneracy. By combining the two best-fitting values of $\rho$ for these two distributions,
any error present in the mean level spacing $\varDelta$ of the Dirac spectrum,
which would result in shifting the unfolded data of $2k-1^{\rm th}$ and $2k^{\rm th}$ eigenvalues
to the same direction, is expected to be cancelled. 
We have confirmed this by generating $10^5$ samples of 
crossover random matrix ensembles with $N=N'=6^4$ and various $\rho\leq 0.50$ and 
by fitting histograms of first four eigenvalues to the analytic results.
Combined values of $\rho$ from these four fittings have reproduced the true input values
within a few per mil of systematic error
(max.~0.5\%), an order of magnitude closer to the input values than using any single individual distribution.
Such an accuracy could neither be hoped for had we used the spectral density $R_1(x)$ for fitting.

\section{Effective theory and low-energy constants}
The Dirac operator $D\!\!\!\!\!/$ of a QCD-like theory with  
quarks in a pseudoreal (real) representation,
such as the fundamental of Sp$(2N)$ (SO$(N)$),
possesses an antiunitary symmetry unlike
QCD with quarks in a complex representation \cite{Ver94}:
$D\!\!\!\!\!/$ commutes with ${\cal C}Z \ast$ (${\cal C}\ast$),
with ${\cal C}$ being the charge conjugation and $\ast$ the complex conjugation.
As $({\cal C}Z\ast)^2=+1$ ($({\cal C}\ast)^2=-1$), $D$ can be brought to a real symmetric
(quaternion selfdual) matrix by a similarity transformation.
Due to this property, the distinction between
left-handed quarks and conjugated right-handed quarks is lost,
leading to the Pauli-G\"{u}rsey extension of the flavor symmetry from ${\rm SU}(N_F)_L\times {\rm SU}(N_F)_R$
to ${\rm SU}(2N_F)=:G$ and its vector subgroup 
from ${\rm SU}(N_F)_V$ to ${\rm Sp}(2N_F)$ or ${\rm SO}(2N_F)=:H$.
Accordingly its low-energy effective theory becomes a nonlinear 
$\sigma$ model on an exotic Nambu-Goldstone manifold $G/H$.

Since the Dirac operator charged under the ${\rm U(1)}$ gauge field is complex,
coupling QCD-like theories with electromagnetism or even 
subjecting them to the constant ${\rm U(1)}$ background
breaks the antiunitary symmetry of $D\!\!\!\!\!/$ and the Pauli-G\"{u}rsey extended flavor symmetry.
In the latter case that is equivalent to putting on a weak imaginary chemical potential $\mu=i \mu_I$,
its effect on the low-energy Lagrangian is systematically incorporated by the
flavor covariantization of the derivatives \cite{KSTVZ00}.
Furthermore, if the theory is in a finite volume $V=L^4$ and the
Thouless energy $E_c\simeq {{F^2}/{\Sigma L^2}}$ is much larger than $m$, 
the path integral is dominated by the zero-mode integration (the $\varepsilon$ regime),
\be
Z=\int_{{\rm SU}(2N_F)} \!\!\!\!\!\!\!\!\!\!\!\!\! dU\,\,\,\exp\(
\frac12 V\Sigma m \,{\rm Re}\,{\rm tr}\, \hat{M} U
-V\mu_I^2 F^2\,{\rm tr}\, (\hat{B} U^\dagger\hat{B} U+\hat{B}\hat{B})
\).
\label{Zchiral}
\ee
Here $U$ is an SU($2N_F$) matrix-valued Nambu-Goldstone field,
$\hat{B}=\sigma_3 \otimes \openone_{N_F}$, 
$\hat{M}=i\sigma_2\otimes \openone_{N_F}$\ $\(\sigma_1 \otimes \openone_{N_F}\)$ 
for quarks in a pseudoreal (real) representation.
$\Sigma=\<\bar{\psi}\psi\>/N_F$ denotes the chiral condensate
and $F$ the pseudo-scalar decay constant,
both measured in the chiral and zero-chemical potential limit.
Note that the above 0D $\sigma$ model for the case of fermions in a real representation 
can as well be derived from the random matrix ensemble (\ref{H}) through the standard procedure: 
(i) introduce $N_F$ species of complex Grassmannian $(N+N')$-vectors $\psi_f, \bar{\psi}_f$
and consider a replicated spectral determinant
$\left<\det (\lambda-H)^{N_F}\right>=
\left<\int d\psi d\bar{\psi}\,\e^{\sum_f \bar{\psi}_f (\lambda-H) \psi_f}\right>$,
where $\left<\cdots\right>$ denotes averaging over $A$ and $B$,
(ii) perform Gaussian integrations over $A$ and $B$,
(iii) introduce a $2N_F\times 2N_F$-matrix valued Hubbard-Stratonovich variable $Q$ and
open up the 4-fermi term,
(iv) perform Gaussian integrations over $\psi$ and $\bar{\psi}$,
(v) take the aforementioned triple-scaling limit and denote the angular part of $Q$
(not fixed by the large-$N$ saddle point equation) as $U$.
Then the coefficients of the mass and chemical-potential terms are identified as
$V\Sigma m=i\pi x$ and $2VF^2\mu_I^2=\pi^2\rho^2$.
By substituting $m\to i \lambda_{\rm Dirac}$ which turns the QCD partition function
into the Dirac spectral determinant, the former equality provides
the definition of unfolded Dirac eigenvalues $x=\lambda_{\rm Dirac}/\varDelta$
due to the Banks-Casher relation $\Sigma=\pi/\varDelta V$.
The latter equality is used to determine $F^2$ from the slope of the $\mu_I$-$\rho$ plot.

\section{Fitting Dirac spectra of SU(2)$\times$U(1) gauge theory}
As the aim of this work is
to demonstrate the validity and advantage of the method and not to approach the continuum, chiral, or thermodynamic limit,
we chose the simplest possible setting on the lattice side:
(i) generate $10^4$ samples of quenched SU(2)=Sp(2) gauge fields $U_\mu(x)$ on an (intentionally)
small lattice $V=6^4$,
with a plaquette action at $\beta_{{\rm SU(2)}}=6/g_{\rm SU(2)}^2=0\sim 1.75$ (step .25), using
the standard heat-bath/overrelaxation algorithm.
(ii-a) multiply the ${\rm SU(2)}$ fields on temporal links $U_0(x)$ by a constant phase $\e^{i\mu_I}$
with $\mu_I=0.00524\sim.05240$ (step .00524), or
(ii-b) generate quenched noncompact ${\rm U(1)}$ gauge fields $A_\mu(x)$ under the Coulomb gauge-fixing condition
\cite{BDHIY07} and multiply the ${\rm SU(2)}$ fields $U_\mu(x)$
by $\exp({ie_{{\rm U(1)}} A_\mu(x)})$, with
$e_{{\rm U(1)}}=0.0004\sim.0024$ (step .0004),
(iii) substitute the gauge fields into an unimproved staggered Dirac operator and diagonalize.

Due to the absence of the ${\cal C}$ matrix, the antiunitary symmetries of staggered Dirac operators 
are swapped between real and pseudoreal representations \cite{HV95}. 
Accordingly, our case with SU(2)$\times$U(1) fundamental fermions indeed 
corresponds to the chGSE-chGUE crossover (\ref{H}).
The low-energy constants are determined by the following steps: 
(I) fit the histogram of each of the two smallest Dirac eigenvalues (i.e.~four counting the Kramers degeneracy) of
the pure SU(2) case to the rescaled chGSE ($\rho=0$) prediction $p_k(\lambda_k/\varDelta)/\varDelta$
by varying $\varDelta$,
(II) combine two optimal values of $\varDelta$ and their variances to determine $\bar{\varDelta}$
and thus $\Sigma=\pi/\bar{\varDelta} V$,
(III) fit the histogram of each of the four smallest {\em unfolded} Dirac eigenvalues $x_k=\lambda_k/\bar{\varDelta}$
of (a) SU(2)$+\mu_I$ or (b) SU(2)$\times$U(1) case 
to the chGSE-chGUE prediction $p_k(x_k)$ by varying $\rho$,
(IV) combine four optimal values of $\rho$ and their variances to determine $\bar{\rho}$ and thus 
$F^2\mu_I^2=(\pi^2/2)\bar{\rho}^2/V$.

We first observe that the four values of $\varDelta$ obtained in the step (I) are mutually consistent,
giving rise to combined relative errors in $\Sigma$ that are
extremely small, $\sim 0.1\%$ (Table 1, top).
One-parameter fittings in the steps (I), (III-a), or (III-b) are quite satisfactory, 
with $\chi^2/{\rm dof}=0.5\sim 1.5$ for all range of parameters in concern (exemplified in Fig.~2, above).
We also confirmed our expectation that the best-fitting values of $\rho$ for $k=1, 3$ and 
those for $k=2, 4$ have a tendency to counter-move,
in favor of cancelling the unfolding ambiguity due to a tiny error within $\varDelta$. 
Relative errors in $\bar{\rho}$ are considerably reduced by the combined use of
four individual eigenvalue distributions (Fig.~2 below), and are no larger than $\pm .018$(stat)$\pm .005$(sys).
Linear response of $\bar{\rho}$ on $\mu_I$ or $e_{\rm U(1)}$ is confirmed
for the SU(2)$+\mu_I$ case (Fig.~3, left), and 
the pseudo-scalar decay constant $F^2$ at various values of $\beta_{\rm SU(2)}$ is obtained from the slopes
(Table 1, middle).
For the SU(2)$\times$U(1) case,
the coefficients (equivalent of $F^2\mu_I^2$) of the ${\rm tr}\,\hat{B} U^\dagger\hat{B} U$ term in
(\ref{Zchiral}) divided by $e_{\rm U(1)}^2$,
extrapolated to $e_{\rm U(1)}\to 0$ 
are summarized in Table 1, bottom.
Complete lattice results, and details of analytic and numerical computations presented
in \S {\bf 2} and \S {\bf 3} will be reported in a subsequent publication.

\begin{table}[h]
$\!\!\!\!\!\!\!\!$
\begin{tabular}{l|llllllll}
$\beta_{\rm SU(2)}$ & 0 & 0.25 & 0.50 & 0.75 & 1.00 & 1.25 & 1.50 & 1.75 \\ 
\hline 
$\Sigma\,a^3$
& 1.310(2) & 1.255(2) & 1.199(1) & 1.139(1) & 1.070(1) & .987(1) & .883(1) & .743(1)\\
$F^2 a^2$
& .284(2) & .268(2) & .247(2) & .226(2) & .205(1) & .178(1) & .153(1) & .115(1)\\
$F^2\mu_I^2 a^4/e_{\rm U(1)}^2$
 & 220(2) & 198(2) & 186(2) & 163(1) & 145(1) & 123(1) & 99.5(8) & 68.0(6)\\
\end{tabular}
\caption{Chiral condensate $\Sigma$ from quenched SU(2) [top], 
pseudo-scalar decay constant $F^2$ from SU(2)+$\mu_I$ [middle],
and an equivalent of $F^2\mu_I^2$ (divided by $e_{\rm U(1)}^2$) from
SU(2)$\times$U(1) [bottom], all in the lattice unit.}
\end{table}
\begin{figure}[ht] 
$\!\!\!\!\!\!\!\!\!\!$
        \includegraphics[bb=0 0 360 235,width=5.32cm]{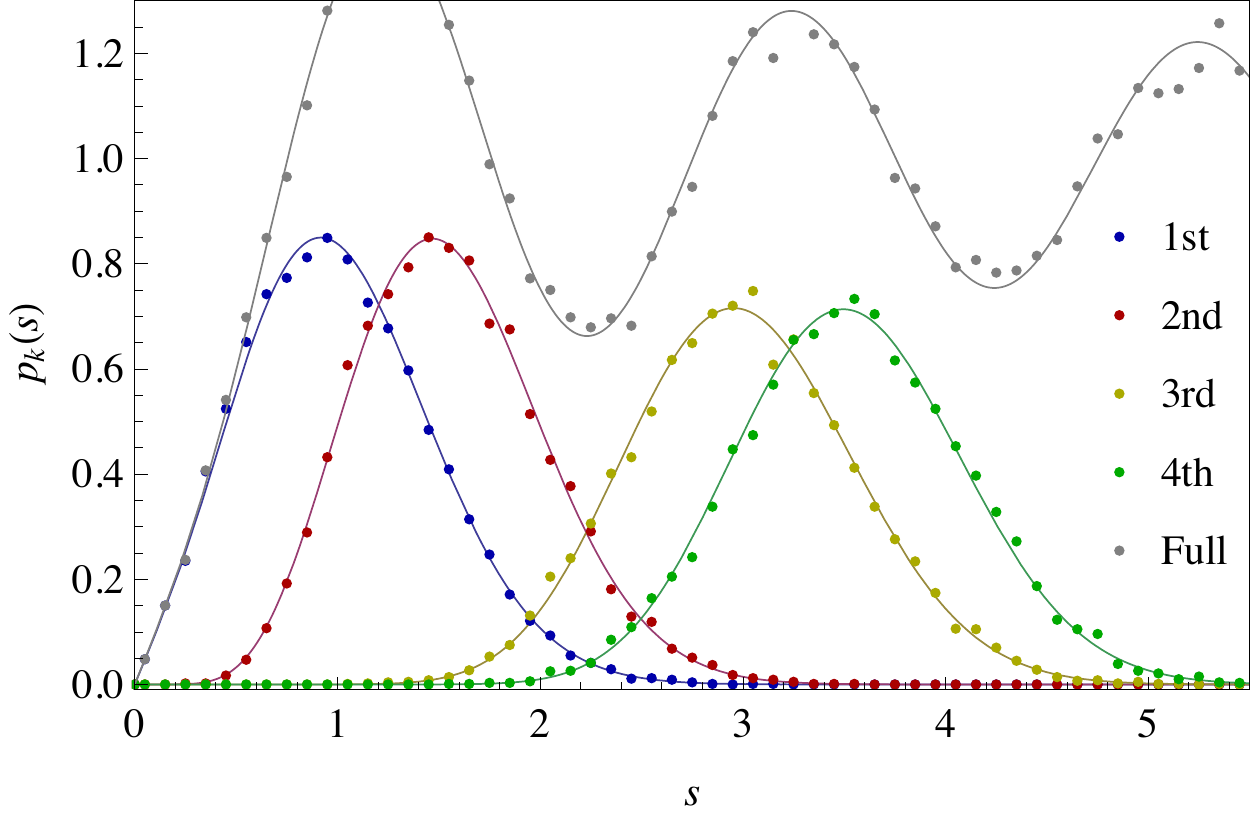}
        \includegraphics[bb=0 0 360 258,width=4.85cm]{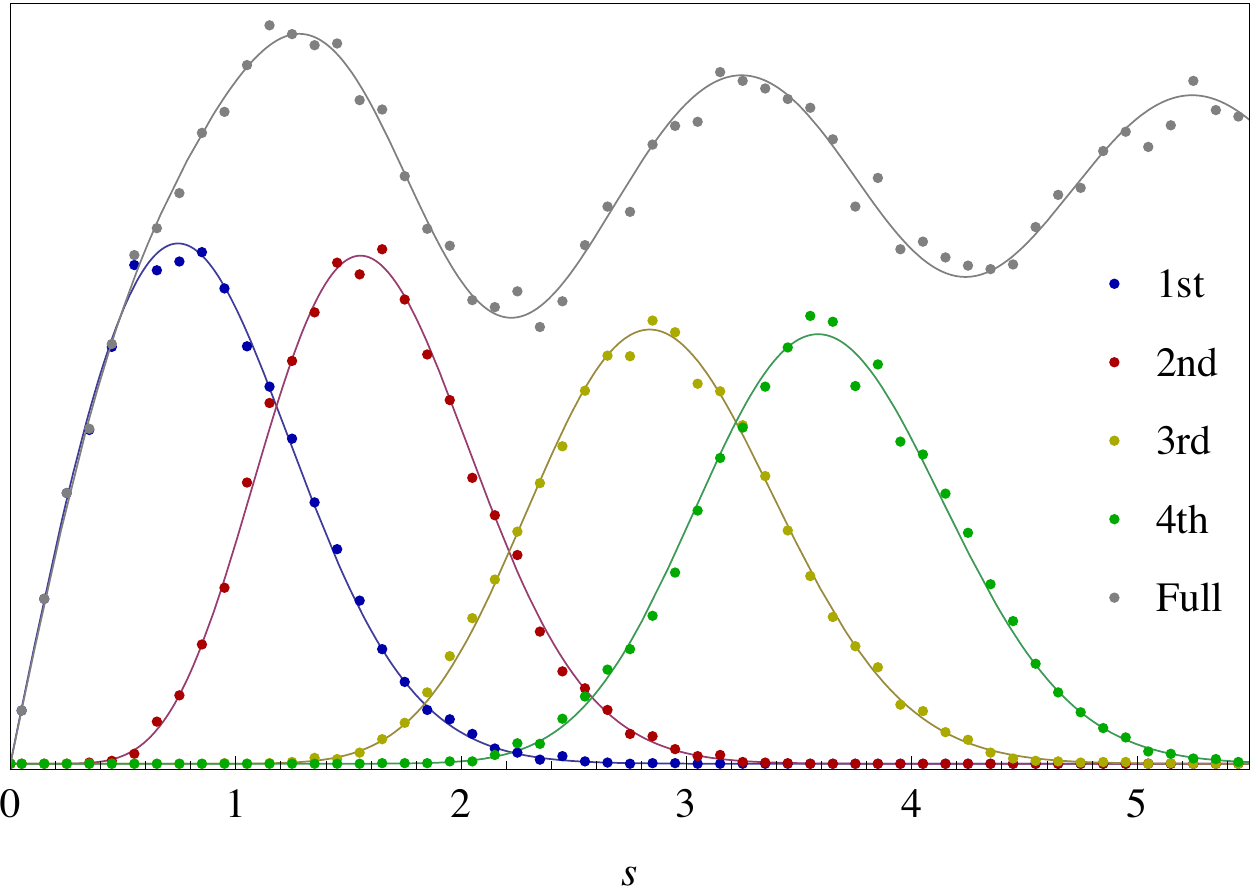}
        \includegraphics[bb=0 0 360 258,width=4.85cm]{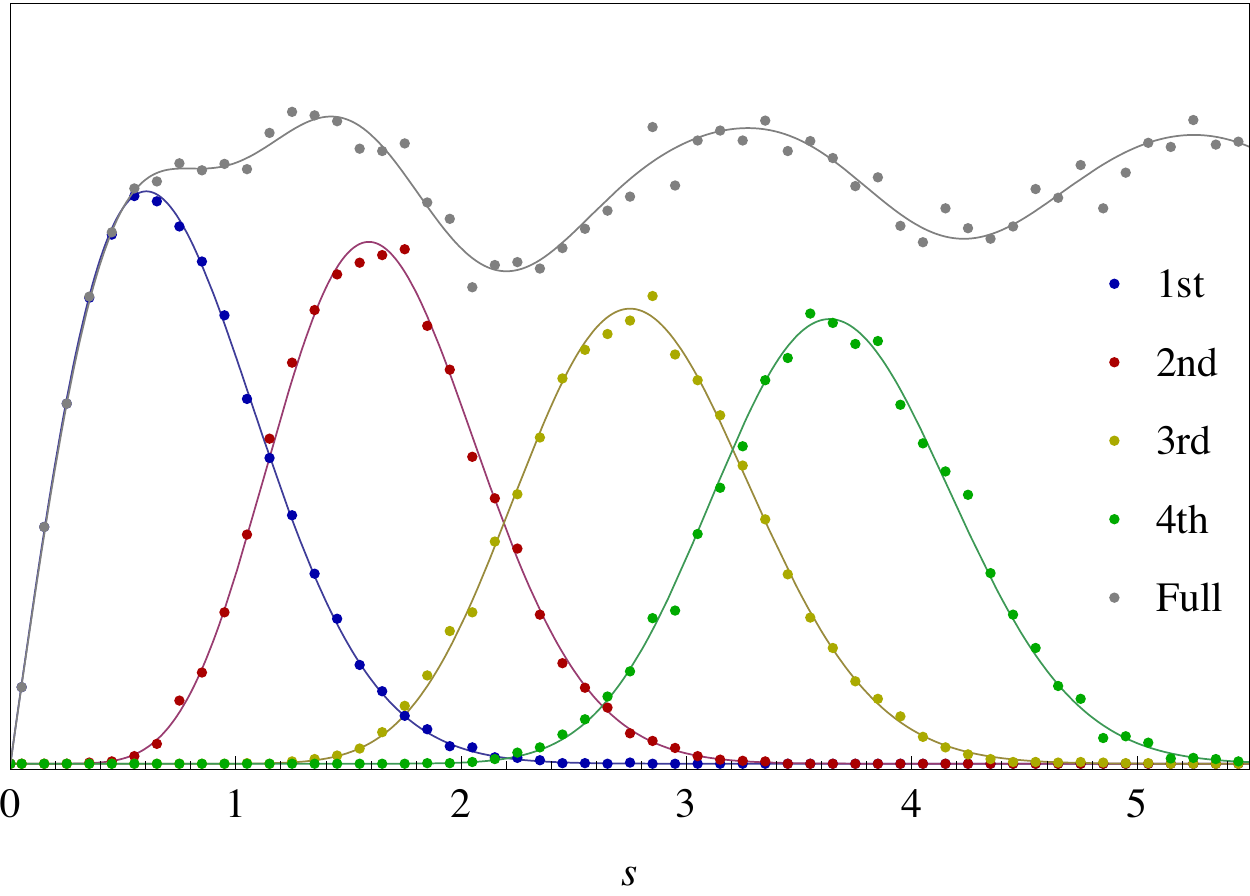}

$\!\!\!\!\!\!\!\!\!\!\!\!$
        \includegraphics[bb=0 0 188 110,width=5.1cm]{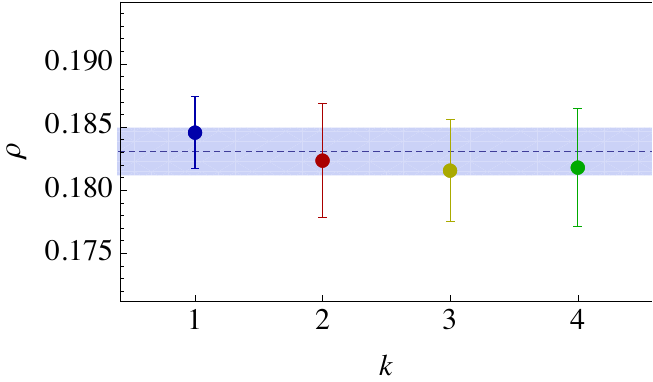}~~
        \includegraphics[bb=0 0 175 110,width=4.8cm]{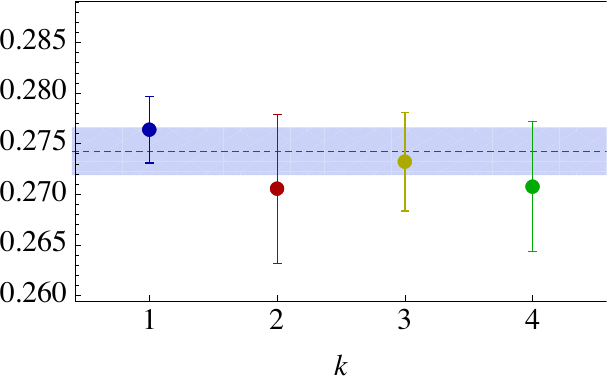}~~
        \includegraphics[bb=0 0 174 109,width=4.8cm]{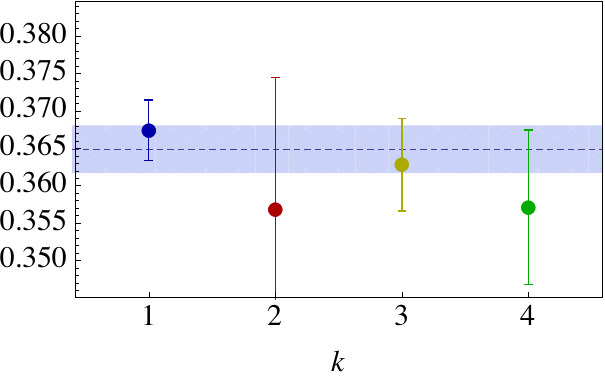}
\caption{Histograms of $k^{\rm th}$ unfolded Dirac eigenvalues for $\beta_{\rm SU(2)}=0.25$ and 
$e_{\rm U(1)}=0.0008, .0012, .0016$
[above, left to right] and best-fitting $p_k(s)\ (k=1\sim4)$ from the chGSE-chGUE crossover. 
The $\rho$ parameter determined for each $k$,
their combined values [dots] and statistical errors [band] are shown below each graph.}
\vspace{5mm}
\centering
\includegraphics[bb=0 0 178 106,width=6cm,clip]{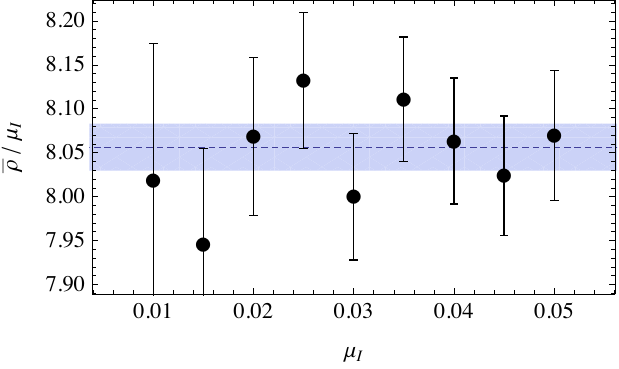}~~~~
\includegraphics[bb=0 0 176 106,width=6cm,clip]{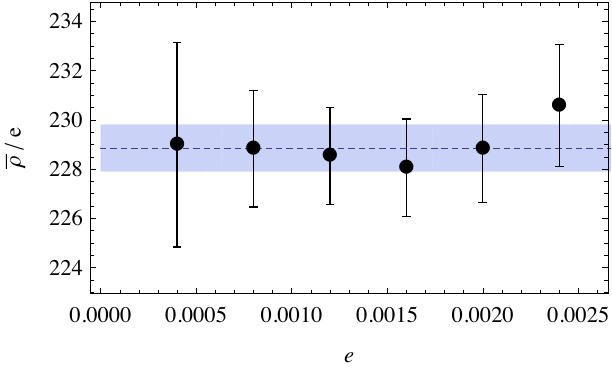}
\caption{Ratios $\bar{\rho}/\mu_I$ for SU(2)+$\mu_I$ at $\beta_{\rm SU(2)}=0.5$ [left] and 
$\bar{\rho}/e_{\rm U(1)}$ for SU(2)$\times$U(1) at $\beta_{\rm SU(2)}=0.25$ [right].
}
\end{figure}

\end{document}